# CONSTRAINTS AND CASIMIRS FOR SUPER POINCARÉ AND SUPERTRANSLATION ALGEBRAS IN VARIOUS DIMENSIONS

BRUNO ZUMINO

*Department of Physics, University of California and Theory Group of Lawrence Berkeley National Laboratory, Berkeley, California 94720*

We describe, for arbitrary dimensions the construction of a covariant and supersymmetric constraint for the massless super Poincaré algebra and we show that the constraint fixes uniquely the representation of the algebra. For the case of finite mass and in the absence of central charges we discuss a similar construction, which generalizes to arbitrary dimensions the concept of the superspin Casimir. Finally we discuss briefly the modifications introduced by central charges, both scalar and tensorial.

## 1. Constraints and Casimirs

Let us start with the super Poincaré algebra,

$$[iJ_{\rho\sigma}, P_\mu] = \eta_{\mu\sigma} P_\rho - (\rho \leftrightarrow \sigma) \tag{1}$$

$$[iJ_{\rho\sigma}, J_{\mu\nu}] = [\eta_{\mu\sigma} J_{\nu\rho} - (\rho \leftrightarrow \sigma)] - (\mu \leftrightarrow \nu) \tag{2}$$

$$[P_\mu, P_\nu] = 0 \quad , \quad [P_\mu, Q] = 0 \tag{3}$$

$$[iJ_{\rho\sigma}, Q] = -\frac{1}{2} \Gamma_{\rho\sigma} Q \tag{4}$$

$$\{Q, \bar{Q}\} = 2i \slashed{P} \tag{5}$$

Here $\eta_{\mu\nu} = \text{diag}(-1, 1, ., 1)$ and $\slashed{P} = \Gamma_\mu P^\mu$, with $\Gamma_\mu$ satisfying the Clifford algebra. Also, $Q$ is a spinor of supercharge, $\bar{Q} = iQ\Gamma^0$ and $\Gamma_{\mu\nu} = \frac{1}{2}[\Gamma_\mu, \Gamma_\nu]$. All spinor indices are suppressed; in particular, $Q^\dagger = (Q^*)^T$ where $Q^*$ is the adjoint of $Q$ and $(\cdot)$ indicates transposition with respect to the spinor indices. Notice also that if $Q$ is a chiral spinor in $D$ spacetime dimensions, then the right-hand side of Eq. (5) should contain a chiral projector $\frac{1}{2}(1 \pm \Gamma)$ and a convenient definition for $\Gamma$ is

$$\Gamma = i^{\frac{1}{2}D-1} \Gamma_0 \Gamma_1 \ldots \Gamma_{D-1}$$

We define the supersymmetry variation of an operator $\mathcal{O}$ to be

$$\delta\mathcal{O} = \{Q, \mathcal{O}\} \quad \text{or} \quad [Q, \mathcal{O}] \tag{6}$$

depending on whether $\mathcal{O}$ is fermionic or bosonic.

Next, we construct two antisymmetric three-tensors, namely

$$W_{\lambda\mu\nu} = P_{<\lambda} J_{\mu\nu>} = \frac{1}{3!} \sum_{perm} \pm P_\lambda J_{\mu\nu} \tag{7}$$

and

$$S_{\lambda\mu\nu} = \bar{Q}\Gamma_{\lambda\mu\nu}Q \tag{8}$$

In four dimensions, Eq. (7) is the dual of the Pauli-Lubanski vector, so that $W$ should be thought of as its generalization to higher dimensions.

Here and in the following, the angular brackets between indices indicate a sum over all permutations of the indices, each taken with a sign and divided by the total number of permutations, as in Eq. (7). Furthermore, $\Gamma_{\lambda\mu\nu} = \Gamma_{<\lambda}\Gamma_\mu\Gamma_{\nu>}$. Using the algebra Eqs. (1–5), it is easy to compute the supersymmetry variation of W. One finds

$$\delta W_{\lambda\mu\nu} = -\frac{i}{2} P_{<\lambda}\Gamma_{\mu\nu>}Q \tag{9}$$

As for S,

$$\delta S_{\lambda\mu\nu} = \delta\bar{Q}\,\Gamma_{\lambda\mu\nu}Q - \bar{Q}\,\Gamma_{\lambda\mu\nu}\,\delta Q \tag{10}$$

The first term in the expression above can be computed using Eq. (5). The second term is different from zero only if $Q$ is a Majorana spinor. If it is a Majorana spinor, $\delta Q$ can again be computed from Eq. (5) using the Majorana condition. The details of the computation vary depending on the spacetime dimension and also, for even dimensions, on whether $Q$ is a chiral or a Dirac spinor. But the final result is that, if the second term is nonzero, then it is exactly equal to the first term. Hence, the supersymmetry variation of $S$ is twice as large when $Q$ is a Majorana spinor. In particular, we find

$$\delta S_{\lambda\mu\nu} = \begin{cases} -2i\,\slashed{P}\Gamma_{\lambda\mu\nu}\,Q & \text{if Q is not a Majorana spinor,} \\ -4i\,\slashed{P}\Gamma_{\lambda\mu\nu}\,Q & \text{if Q is a Majorana spinor} \end{cases} \tag{11}$$

It is convenient to rewrite the variation of S as the sum of two terms, using the identity

$$\slashed{P}\Gamma_{\lambda\mu\nu} = 6P_{<\lambda}\Gamma_{\mu\nu>} - \Gamma_{\lambda\mu\nu}\slashed{P} \tag{12}$$

The first term has the same form as the variation of $W$ in Eq. (9), and can be used to cancel it if we take an appropriate linear combination of $W$ and $S$. The second term, instead, yields a variation proportional to $\not{P}Q$, and $\not{P}Q = 0$ in a massless representation. We denote the relative coefficient between $W$ and $S$ by $\kappa$ and their linear combination by $\Delta$,

$$\Delta_{\lambda\mu\nu} \equiv W_{\lambda\mu\nu} - \kappa S_{\lambda\mu\nu} \tag{13}$$

It should be clear from the discussion above that the value of $\kappa$ depends only on whether $Q$ is or is not a Majorna spinor and, in particular,

$$\kappa = \begin{cases} \frac{1}{24} & \text{if Q is not a Majorana spinor,} \\ \frac{1}{48} & \text{if Q is a Majorana spinor} \end{cases} \tag{14}$$

With the value of $\kappa$ as above, we find

$$\delta \Delta_{\lambda\mu\nu} = -\frac{i}{12} \Gamma_{\lambda\mu\nu} \not{P} Q \tag{15}$$

so that is possible to impose the constraints

$$P^2 = 0 \ , \ \not{P}Q = 0 \ , \ \Delta_{\lambda\mu\nu} = 0 \tag{16}$$

consistently with the full super Poincaré algebra. Actually, to impose consistently Eq. (16), one also needs $\delta(\not{P}Q) \propto P^2$, which is true.

The constraints Eq. (16) were found in 11-dimensional spacetime in the course of the off-shell quantization of the superparticle [1], with the appropriate value $\kappa = 1/48$ for the relevant coefficient ($Q$ is a Majorana spinor in 11 dimensions). The constraint $\Delta$ actually fixes completely the representation, and in 11 dimensions fixes it to be the supergravity multiplet. In the case of extended supersymmetry one can construct a tensor analogous to $\Delta_{\lambda\mu\nu}$ in a straightforward way. Namely, if there are $N$ supercharges $Q_I$, then

$$\Delta_{\lambda\mu\nu} \equiv W_{\lambda\mu\nu} - \sum I = 1^N \kappa_I \bar{Q}_I \Gamma_{\lambda\mu\nu} Q_I \tag{17}$$

where each of the $\kappa_I$ is given by Eq. (14). Then

$$\delta_I \Delta_{\lambda\mu\nu} \equiv [Q_I, \Delta_{\lambda\mu\nu}] = -\frac{i}{12} \Gamma_{\lambda\mu\nu} \tag{18}$$

and again $\Delta$ can be set to zero consistently. In four spacetime dimensions the $\Delta$ tensor is only one of a continuous class of supercovariant objects that can be constructed. Indeed, if one defines

$$\Delta^{(\chi)}_{\lambda\mu\nu} \equiv \Delta_{\lambda\mu\nu} - \frac{1}{3} \chi P^\alpha \varepsilon_{\alpha\lambda\mu\nu} \tag{19}$$

where $\chi$ is an arbitrary real number and $\varepsilon$ is the completely antisymmetric tensor with $\varepsilon_{0123} = +1$, then the supersymmetry variation of $\Delta^{(\chi)}$ is the same as that of $\Delta$. Hence $\Delta^{(\chi)} = 0$ is also a good constraint, compatible with the full super Poincaré algebra. We will elaborate on this point in the next section.

It is instructive to compare the $\Delta$ tensor to a similar construction which is useful for massive representations. We could rewrite the variation Eq. (11) of $S$ using the identity

$$\slashed{P}\Gamma_{\lambda\mu\nu} = 3P_{<\lambda}\Gamma_{\mu\nu>} + \frac{1}{2}[\slashed{P}, \Gamma_{\lambda\mu\nu}] \qquad (20)$$

instead of (12). Again the first term can be used to cancel the variation of $W$ if a suitable relative coefficient between $W$ and $S$ is chosen. Let us emphasize that the coefficient needed differs from $\kappa$ by a factor of 2. We call the linear combination $C_{\lambda\mu\nu}$ and the relative coefficient $\rho$. Hence,

$$C_{\lambda\mu\nu} \equiv W_{\lambda\mu\nu} - \rho S_{\lambda\mu\nu} \qquad (21)$$

with

$$\rho = \begin{cases} \frac{1}{12} & \text{if Q is not a Majorana spinor,} \\ \frac{1}{24} & \text{if Q is a Majorana spinor} \end{cases} \qquad (22)$$

Then

$$\delta C_{\lambda\mu\nu} = -\frac{i}{12}[\Gamma_{\lambda\mu\nu}, \slashed{P}]Q \qquad (23)$$

and because of the identity $[P^\lambda \Gamma_{\lambda\mu\nu}, \slashed{P}] = 0$, the antisymmetric tensor

$$C_{\mu\nu} \equiv P^\lambda C_{\lambda\mu\nu} \qquad (24)$$

is invariant under supersymmetry transformations. Then the scalar

$$C \equiv C_{\mu\nu}C^{\mu\nu} \qquad (25)$$

is a Casimir of the full super Poincaré algebra and can be used to label its massive representations. For massless representations, on the other hand, it is possible to show that $C$ vanishes identically. In that case $\Delta$ is a more useful quantity to consider. Notice that $C$ generalizes to arbitrary dimension a four-dimensional Casimir constructed in [2-5], where the eigenvalues of that Casimir were termed "superspin".

## 2. Meaning of the Constraint $\Delta = 0$

To investigate how $\Delta = 0$ constrains massless representations of the super Poincaré algebra, we choose a frame in which $P = (E, E, 0, \ldots, 0)$ (light-cone frame). In D spacetime dimensions, this choice breaks $SO(D-1,1)$ down to the 'little group' $ISO(D-2)$, namely the group of rotations and translations in $D-2$ dimensions. For convenience, we introduce Latin indices of two types, $a, b, c = 0, 1$ and $i, j, k = 2, \ldots, D-1$, so that we can express the choice of frame with

$$P^a = E \ , \ P^i = 0 \tag{26}$$

Then the components of $W$ are as follows:

$$W_{abc} = W_{ijk} = 0 \tag{27}$$

$$W_{abi} = \varepsilon_{ab} \frac{E}{3}(J_{i0} - J_{i1}) \equiv \varepsilon_{ab} \frac{E}{3} A_i \tag{28}$$

$$W_{aij} = \pm \frac{E}{3} J_{ij} \tag{29}$$

where $\varepsilon_{ab}$ is the antisymmetric tensor in two dimensions with $\varepsilon_{01} = +1$ and the upper sign in Eq. (29) holds when $a = 1$, the lower when $a = 0$; similarly in Eqs. (34) and (37) below. Note that $A_i = (J_{i0} - J_{i1})$ are precisely the generators of the translations of $ISO(D-2)$.

Before evaluating the components of $S$, we need to discuss how the frame choice Eq. (26) affects the supercharges $Q$ which, in a massless representation, are subject to the constraint $\not{P}Q = 0$. The answer is that some components are projected out. Indeed

$$\Pi_+ Q = 0 \ , \ \Pi_- Q = 0 \tag{30}$$

where $\Pi_+$ and $\Pi_-$ are complementary projectors given by

$$\Pi_\pm = \frac{1}{2}(1 \pm \Gamma^1 \Gamma^0) \tag{31}$$

Using Eq. (30) and performing some algebra, we see that the components of $S$ are

$$S_{abc} = S_{ijk} = 0 \tag{32}$$

$$S_{abi} = 0 \tag{33}$$

$$Saij = \mp i Q^\dagger \Gamma_{ij} Q \tag{34}$$

Therefore, the components of $\Delta$ are

$$\Delta_{abc} = \Delta_{ijk} = 0 \tag{35}$$

$$\Delta_{abi} = \varepsilon_{ab} \frac{E}{3} A_i \tag{36}$$

$$\Delta_{aij} = \pm \left( \frac{E}{3} J_{ij} + i\kappa Q^\dagger \Gamma_{ij} Q \right) \tag{37}$$

with $\kappa$ given by Eq. (14). We see that setting $\Delta = 0$ is equivalent to imposing the pair of conditions

$$A_i = 0 \tag{38}$$

$$J_{ij} = -3i \frac{\kappa}{E} Q^\dagger \Gamma_{ij} Q \tag{39}$$

The first condition requires that the translations of the little group be represented trivially. This is desirable on physical grounds since a nontrivial representation would lead to unwanted continuous degrees of freedom, by a standard field theory argument. The second condition, on the other hand, restricts the eigenvalues of $J_{ij}$ to be those of the quadratic operator to the right of Eq. (39). Those eigenvalues can be computed explicitly in any dimension. They are of course independent of the values of $i$ and $j$, because the frame choice Eq. (26) does not break the rotational invariance in the $i$ and $j$ indices. The eigenvalues are quantized as a result of the supersymmetry algebra Eq. (5). In this frame, the algebra can also be written as

$$\{Q, Q^\dagger\} = 4E\Pi_+ \tag{40}$$

from which it follows that the nonzero components of $Q$ are proportional to fermionic oscillators. How many oscillators exactly will depend on the spacetime dimension and on what kind of spinor $Q$ is (for instance, a chiral or a Majorana condition will each reduce by half the number of independent oscillators). In the end, the right-hand side of Eq. (39) can be written as a simple function of several fermionic number operators. Hence, the eigenvalues of $J$ and their multiplicities can be easily computed and from that a representation can be inferred uniquely.

Two remarks are in order. When extended supersymmetry is present all that we have done can be repeated with only minor changes. The principal difference is that Eq. (39) is replaced by

$$J_{ij} = -3i \sum_{I=1}^{N} \frac{\kappa_I}{E} Q_I^\dagger \Gamma_{ij} Q_I \tag{41}$$

which in turn can be written as a function of $N$ sets of number operators. The second remark concerns the case of four spacetime dimensions, where a continuous class of constraints $\Delta^{(\chi)}$ exists, as mentioned in the previous section. In four dimensions the little group is $ISO(2)$ and it consists of the helicity and of two translations. With our choice of frame Eq. (26), the generators are $J_{23}$, $A_2$ and $A_3$, respectively. Now, setting $\Delta^{(\chi)} = 0$ adds a shift to the eigenvalues of $J_{23}$, namely, to the helicities of the representation, while, interestingly, the constraints $A_2 = 0$ and $A_3 = 0$ are unaffected,

$$A_2 = A_3 = 0 \ , \ J_{23} = -3i \frac{\kappa}{E} Q^\dagger \Gamma_{23} Q + \chi \tag{42}$$

In summary, all possible representations with $A_2 = A_3 = 0$ are recovered as $\chi$ varies. It should not come as a surprise that $\chi$ appears to be a continuous variable, because our construction is purely algebraic, whereas the quantization of the helicities in four dimensions is a consequence of the topology of the little group.

## 3. The Case of Finite Mass

For completeness, we present in this section a discussion of the tensor $C_{\mu\nu}$ defined in Eq. (24). We proceed along the lines of the discussion of $\Delta$. What follows is a generalization to generic spacetime dimensions of similar arguments that can be found in [2-5] for the four-dimensional case.

To begin, we choose a frame in which $P = (m0, ., , , 0)$ (rest frame). The little group is $SO(D-1)$ and it is generated by $J_{ij}$ where $i, j = 1, \ldots, D1$. In the rest frame, $\slashed{P} = m\Gamma_0$, and the supersymmetry algebra Eq. (5) becomes $\{Q, Q^\dagger\} = 2m$. If we rescale the supercharge $Q$ by defining $a = Q/\sqrt{2m}$, then $a$ satisfies

$$\{a, a^\dagger\} = 1 \tag{43}$$

$$[iJ_{ij}, a] = -\frac{1}{2}\Gamma_{ij}a \tag{44}$$

$$[iJ_{ij}, a^\dagger] = +\frac{1}{2}\Gamma_{ij}a^\dagger \tag{45}$$

where the latter two equations follow from Eq. (4). In the rest frame, the components of $C_{\mu\nu}$ are

$$C_{0\mu} = 0 \quad , \quad C_{ij} = -\frac{m^2}{3}\left[J_{ij} + 6i\rho a^\dagger \Gamma_{ij} a\right] \tag{46}$$

with $\rho$ given by Eq. (22).

We now define the tensors

$$T_{ij} = -6i\rho a^\dagger \Gamma_{ij} a \tag{47}$$

and

$$Y_{ij} = -3C_{ij}/m^2 \tag{48}$$

The point of these definitions is that $T_{ij}$ and $Y_{ij}$ are angular momentum operators, in the sense that they satisfy each the commutation relations of the generators of the little group $SO(D-1)$, exactly as $J_{ij}$ does. Furthermore, $T$ and $Y$ commute with one another. Equation (46) becomes

$$J_{ij} = Y_{ij} + T_{ij} \tag{49}$$

and therefore we can conclude that $J$ is the composition of two independent angular momentum operators $T$ and $Y$.

## 4. Concluding Remarks

We introduced a covariant tensor $\Delta$ which, in the case of a massless representation of the super Poincaré algebra, is also supersymmetric. Imposing $\Delta = 0$ is a supervcovariant way to fix the representation completely, including the generators of the translations in the little group. In particular, the translations are represented trivially, as required on physical grounds.

For the case of nonvanishing mass, we have constructed angular momentum operators $Y_{ij}$ and a Casimir $C$. The latter generalizes to higher dimensions the superspin operator. We have also shown how $Y_{ij}$ can be used to construct representations of the super Poincaré algebra.

So far we have ignored the possibility that some of the generators of the algebra are central charges. In the case of non-vanishing mass, when scalar central charges are present, the results described above can be generalized for any space-time dimension. One can keep the same definition of $\Delta$ as for the massless case. Since the algebra is now different (see e.g. ref. [6], pages 390-393) the vanishing of $\Delta$ picks up representations which resemble those for zero mass, the mass term in the r.h.s. of the algebra being compensated by the terms involving the scalar central charges in the case of BPS states. These BPS representations correspond to one-half the supersymmetry of the massive algebra with no central charges. It is well known (see ref. [7] for the case of four space-time dimensions) that there are other BPS representations, in which more than one-half supersymmetry is preserved, Again, $\Delta = 0$ picks out very special and physically interesting representations. Finally, one can extend the results of the previous sections to the case of nonscalar, tensorial central charges [8, 9, 10], which occur in presence of extended objects in space-time: membranes, domain walls etc., (clearly tensorial central charges restrict the super Poincaré algebra to a super translation algebra, see e.g. ref. [6], pages 397-401, where basic properties of tensorial central charges are discussed; on pages 408, 409 useful references are given). It is interesting that, again, the case of one-half supersymmetry emerges as interesting, for instance in the case of the supermembrane in eleven dimensional space-time [8].

**Acknowledgments**

I would like to thank Itzhak Bars, Mary K. Gaillard and Andrea Pasqua for useful discussions. This work was supported in part by the Director, Office of Science, Office of High Energy and Nuclear Physics, of the U.S. Department of Energy under Contract No. DE-AC0376SF00098, and in part by the NSF under Grant No. 22386-13067.